\documentclass[conference,a4paper]{IEEEtran}
\IEEEoverridecommandlockouts

\usepackage{cite}
\usepackage{amsmath,amssymb,amsfonts}
\usepackage{algorithm}
\usepackage{algpseudocode}
\usepackage{booktabs}
\usepackage{array,tabularx}
\usepackage{graphicx}
\usepackage{xcolor}
\usepackage{textcomp}
\usepackage{xspace}
\usepackage{url}
\usepackage{listings}
\usepackage{tikz}
\usetikzlibrary{arrows.meta,positioning,calc,shapes.geometric,fit}

\definecolor{tcgblue}  {RGB}{31, 119, 180}   
\definecolor{tcggreen} {RGB}{ 44, 160,  44}  
\definecolor{tcgorange}{RGB}{255, 127,  14}  
\definecolor{tcgred}   {RGB}{214,  39,  40}  
\definecolor{tcgyellow}{RGB}{255, 187, 120}  
\definecolor{tcggray}  {RGB}{127, 127, 127}  
\definecolor{tcglight} {RGB}{231, 231, 240}  

\tikzset{
  block/.style   = {rectangle, draw, rounded corners=2pt, align=center,
                    inner sep=4pt, minimum height=18pt, font=\footnotesize},
  store/.style   = {cylinder, shape border rotate=90, draw, aspect=0.25,
                    align=center, inner sep=2pt, minimum height=18pt,
                    minimum width=42pt, font=\footnotesize,
                    fill=tcgyellow!60, draw=tcgyellow!90!black},
  actor/.style   = {block, fill=tcglight, draw=tcgblue!60!black},
  trust/.style   = {block, fill=tcgblue!85, text=white,
                    draw=tcgblue!60!black, font=\footnotesize\bfseries},
  allowx/.style  = {block, fill=tcggreen!20, draw=tcggreen!75!black},
  reviewx/.style = {block, fill=tcgorange!25, draw=tcgorange!85!black,
                    font=\footnotesize\itshape},
  blockx/.style  = {block, fill=tcgred!25, draw=tcgred!75!black,
                    font=\footnotesize\itshape},
  shadowx/.style = {block, fill=tcggray!15, draw=tcggray,
                    font=\footnotesize\itshape},
  auditx/.style  = {block, fill=tcglight, draw=tcgblue!40!black},
  flowallow/.style  = {-{Stealth[length=4pt,width=4pt]}, thick, tcggreen!75!black},
  flowreview/.style = {-{Stealth[length=4pt,width=4pt]}, thick, tcgorange!85!black},
  flowblock/.style  = {-{Stealth[length=4pt,width=4pt]}, thick, tcgred!75!black},
  flow/.style       = {-{Stealth[length=4pt,width=4pt]}, thick, tcgblue!60!black},
  flowdash/.style   = {-{Stealth[length=4pt,width=4pt]}, dashed, thick, tcggray},
  lifeline/.style = {thick, densely dashed, tcgblue!40!black},
  msg/.style      = {-{Stealth[length=4pt,width=4pt]}, semithick, tcgblue!70!black,
                     font=\fontsize{7}{8}\selectfont},
  msgret/.style   = {-{Stealth[length=4pt,width=4pt]}, semithick, dashed,
                     tcgblue!50!black, font=\fontsize{7}{8}\selectfont},
  rolebox/.style  = {block, fill=tcgblue!85, text=white,
                     draw=tcgblue!60!black, minimum width=44pt,
                     font=\fontsize{7}{8}\selectfont\bfseries}
}

\def\BibTeX{{\rm B\kern-.05em{\sc i\kern-.025em b}\kern-.08em
    T\kern-.1667em\lower.7ex\hbox{E}\kern-.125emX}}

\newcommand{\tcg}{ToolContractGate\xspace}
\newcommand{\textsmaller}[1]{{\fontsize{8}{9}\selectfont #1}}

\AddToHookNext{shipout/foreground}{%
  \begin{tikzpicture}[remember picture,overlay]
    \node[anchor=south,inner sep=0pt] at ([yshift=32pt]current page.south) {%
      \begin{minipage}{\textwidth}
      \footnotesize
      \textbf{Author-accepted manuscript.} Accepted for publication in the 2026 International Conference on Computer and Applications (ICCA).\par\smallskip
      \textcopyright\ 2026 IEEE. Personal use of this material is permitted. Permission from IEEE must be obtained for all other uses, in any current or future media, including reprinting/republishing this material for advertising or promotional purposes, creating new collective works, for resale or redistribution to servers or lists, or reuse of any copyrighted component of this work in other works.
      \end{minipage}%
    };
  \end{tikzpicture}%
}

\begin{document}

\title{\tcg: Versioned Trusted Contracts for Drift-Safe Agent Tool Invocation}

\author{\IEEEauthorblockN{Sachin Gupta}
\IEEEauthorblockA{\textit{Independent Researcher}\\
San Jose, California, USA\\
sachinkg12@gmail.com}
}

\maketitle
\thispagestyle{empty}

\begin{abstract}
Agent tool approval and input validation do not, by themselves, bind a request to an approved source artifact. ToolContractGate provides a deterministic runtime contract evaluator for OpenAPI-derived agent tools. Each versioned ToolContract combines a stored OpenAPI fingerprint with access, schema, and behavioral declarations. The full evaluator applies ten checks, including a cached boot-time provenance result; four execution modes and five evaluation configurations support baseline, ablation, and non-blocking shadow analysis. On a synthetic benchmark of 45 cases (35 unsafe cases across 12 drift classes and 10 safe cases), the full pipeline detects all 35 faults: 29 BLOCK, 3 REVIEW, and 3 WARN decisions, with no false blocks on the safe cases. The archived run reports a mean gate latency of 0.18\,ms and a 95th percentile (p95) of 2\,ms using integer-millisecond measurements. The open-source artifact includes seven OpenAPI specifications, twelve enterprise contracts, 29 unit tests, and a 900-record evaluation. Results characterize contract-decision coverage; downstream enforcement requires a host that honors the returned decision. Production traffic, semantic backend drift, and dynamic provenance refresh remain evaluation and integration opportunities.
\end{abstract}


\section{Introduction}

Large language model (LLM) agents increasingly invoke external tools through structured interfaces such as the Model Context Protocol (MCP)~\cite{mcp_spec}. An operator may approve a tool name while the surrounding API, access requirements, or declared side effects evolve. Approval of a name alone does not establish that a subsequent request is consistent with the operator's approved specification and policy.

Three failure classes motivate this work. First, \emph{structural drift}: a request uses an obsolete version or invalid enum value, or omits a required field. Second, \emph{misbinding}: a request references a contract for \texttt{ticket.create} but names \texttt{ticket.close}. Third, \emph{declared escalation}: the request reports stronger side effects or weaker confirmation than its contract permits. These are observable contract discrepancies. A backend that changes behavior while preserving all visible declarations is a separate semantic-drift problem.

We propose a \emph{trusted runtime contract}: a versioned ToolContract associated with an approved OpenAPI artifact and evaluated for each request. Its fingerprint links the contract to the approved file bytes verified at startup; typed predicates compare the supplied request with contract policy. An integrating host can then allow, warn, hold for review, or suppress the invocation. Our contributions are:
\begin{itemize}
  \item A contract model combining version compatibility, tool binding, scopes, input schema, side effects, confirmation, payload limits, audit class, and OpenAPI provenance, with configurable violation severity.
  \item A deterministic ten-check evaluator with documented severity precedence, four execution modes, and configurable ablations, including a shadow mode that preserves would-be decisions while always returning runtime permission.
  \item A boot-time SHA-256 verifier and a reproducible 45-case benchmark evaluated under four modes and five configurations (900 evaluations), with open code, contracts, tests, and recorded results.
\end{itemize}

\section{Related Work}
\label{sec:related}

\textbf{Host policies and protocol metadata.} Host-side approval and tool filtering decide whether a call should proceed. MCP defines \texttt{ToolAnnotations}, including read-only, destructive, idempotent, and open-world hints, but these declarations are not guarantees and should not be trusted merely because a server supplies them~\cite{mcp_spec}. The draft interceptor proposal SEP-1763 describes client/server hooks for validation, transformation, and observation~\cite{mcp_interceptors}. Such hooks can host a contract evaluator; the hook mechanism alone does not define an approved-artifact contract policy. ToolContractGate stores MCP-aligned annotations while its checks use explicit typed side-effect and confirmation fields.

\textbf{Identity and prompt injection.} SPIFFE provides workload identities~\cite{spiffe_spec}; identity establishes who is communicating rather than whether a particular tool request matches an approved contract. Greshake et al.~\cite{greshake2023indirect} demonstrate indirect prompt injection through attacker-controlled content. Contract evaluation complements content defenses but does not reject every malicious intent expressed through an otherwise permitted tool call.

\textbf{Generation and offline governance.} Mastouri et al.~\cite{mastouri2025automcp} study REST-to-MCP wrapping and automated server generation. ToolContractGate addresses the downstream decision point; its released contracts are manually seeded rather than automatically derived. OpenAPI-diff~\cite{openapidiff} compares specification revisions, while Pact~\cite{pact} tests consumer--provider compatibility before deployment. These techniques can supply evidence for contract updates but do not themselves implement this per-call policy. Diffy~\cite{diffy} compares service behavior using shadow testing; our shadow mode instead records contract decisions without changing runtime permission.

\textbf{Risk and provenance frameworks.} NIST AI RMF~\cite{nist_airmf} and OWASP's LLM risk guidance~\cite{owasp_llm_top10} motivate governance, measurement, and controls on excessive agency. Syed et al.~\cite{syed2025crossagent} study provenance-aware prompt-injection defense, a different use of provenance from the artifact-digest linkage here. Panwar and Abdelrahman~\cite{panwar2025review} review agentic-AI cybersecurity. Table~\ref{tab:related} locates the respective control points; these mechanisms are complementary rather than interchangeable.

\begin{table*}[t]
\centering
\caption{Complementary controls and the role of contract evaluation. Entries describe their stated control points rather than an exhaustive comparison of products.}
\label{tab:related}
\small
\setlength{\tabcolsep}{5pt}
\begin{tabularx}{\textwidth}{@{}p{0.23\textwidth} p{0.17\textwidth} X@{}}
\toprule
\textbf{Control} & \textbf{Point of application} & \textbf{Relationship to contract evaluation}\\
\midrule
Host approval and tool filtering & Host, per invocation & Approval controls whether a call proceeds; a contract adds explicit artifact and typed-policy checks.\\
MCP interceptor proposal~\cite{mcp_interceptors} & Client/server hooks & Draft validation and observation mechanism; a contract policy can be implemented at a hook.\\
OpenAPI-diff~\cite{openapidiff}; Pact~\cite{pact} & CI / provider testing & Specification differences and compatibility tests inform approved contract revisions.\\
Diffy~\cite{diffy} & Shadow service testing & Compares service behavior; our shadow mode compares invocation declarations against a contract.\\
SPIFFE~\cite{spiffe_spec} & Workload identity & Establishes caller identity; authorization and contract admission require additional policy.\\
Prompt-injection research~\cite{greshake2023indirect} & Agent input/context & Studies attacks and mitigation limits; contract checks address invocation-policy discrepancies.\\
\midrule
\textbf{ToolContractGate} & \textbf{Per-call evaluator} & Ten checks combine contract policy with cached startup integrity of an approved OpenAPI artifact. Host integration must enforce the result.\\
\bottomrule
\end{tabularx}
\end{table*}

\section{System Model}
\label{sec:model}

\begin{figure}[t]
\centering
\begin{tikzpicture}[font=\footnotesize,node distance=7pt,
 box/.style={draw,rounded corners=2pt,align=center,inner sep=4pt,text width=6.9cm},
 arr/.style={-{Stealth[length=4pt]},semithick}]
\node[box,fill=tcgyellow!35] (art) {Approved OpenAPI bytes + versioned contracts};
\node[box,below=of art,fill=tcgyellow!20] (reg) {Startup SHA-256 verification\\Registry + compiled schema + cached validity};
\node[box,below=of reg,fill=tcgblue!10] (gate) {Request + trusted host metadata\\$\downarrow$\\\textbf{ToolContractGate evaluator}};
\node[box,below=of gate] (d) {Decision + runtime decision + violations};
\node[box,below=12pt of d,dashed,fill=tcglight!50] (host) {\textbf{Required host integration}\\Allow/warn: invoke tool; block: suppress call\\Review: hold for operator; persist audit record};
\draw[arr] (art)--(reg); \draw[arr] (reg)--(gate); \draw[arr] (gate)--(d);
\draw[arr,dashed] (d)--(host);
\end{tikzpicture}
\caption{Implemented contract evaluation and its deployment boundary. SHA-256 verification occurs at startup. The dashed block identifies host responsibilities; the benchmark endpoint returns decisions without implementing downstream execution, review routing, or durable audit storage.}
\label{fig:architecture}
\end{figure}
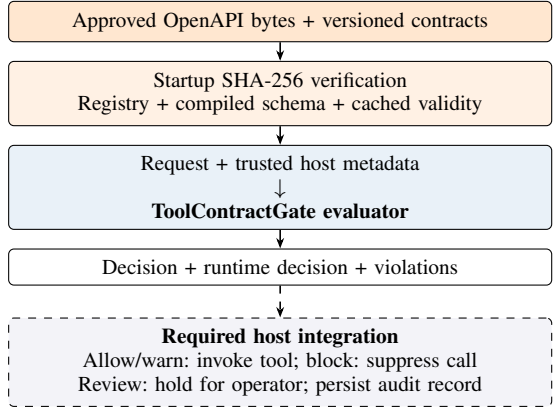

Figure~\ref{fig:architecture} separates the implemented evaluator from the host responsibilities needed for enforcement. The trusted registry holds approved contracts and compiled schemas. A request supplies contract identity/version, tool name, scopes, payload, and optional behavioral metadata. The evaluator returns a structured decision; an enforcing host must obtain trustworthy metadata and honor that decision before forwarding the call.

\subsection{ToolContract}

A ToolContract has nine typed sections (Figure~\ref{fig:contract}). The contract records identity and ownership; required scopes and audit class; limits; behavioral declarations; compatible versions; drift policy; provenance; MCP-aligned annotations; and input schema. The side-effect order is \texttt{read\_only} $<$ \texttt{write} $<$ \texttt{destructive}; audit classes are \texttt{routine} $<$ \texttt{internal} $<$ \texttt{sensitive} $<$ \texttt{regulated}. Payload size is checked against \texttt{max\_payload\_bytes}; \texttt{max\_rps} is metadata without a rate counter in this prototype.

\begin{figure}[t]
\centering
\begin{tikzpicture}[
  yamlf/.style  = {anchor=west, inner sep=0pt,
                   font=\ttfamily\fontsize{6.7}{8}\selectfont},
  yhdr/.style   = {anchor=west, inner sep=0pt,
                   font=\ttfamily\fontsize{6.7}{8}\selectfont\itshape,
                   color=gray!55!black},
  seclbl/.style = {anchor=west, inner sep=1pt,
                   font=\fontsize{7}{8.5}\selectfont\bfseries}
]
\def\codeX{0.12}    
\def\bracX{4.95}    
\def\labX{5.18}     
\def\bw{0.12}       
\def\dy{0.30}       

\node[yhdr] at (\codeX, 0.45) {\# contracts/catalog.search.v1.yaml};

\node[yamlf] at (\codeX, 0)        {id: catalog.search};
\node[yamlf] at (\codeX,  -1*\dy)  {version: v1};
\node[yamlf] at (\codeX,  -2*\dy)  {tool: catalog.search};
\node[yamlf] at (\codeX,  -3*\dy)  {owner: catalog-platform};
\node[yamlf] at (\codeX,  -4*\dy)  {audit\_class: internal};
\node[yamlf] at (\codeX,  -5*\dy)  {required\_scopes: [catalog.read]};
\node[yamlf] at (\codeX,  -6*\dy)  {limits:};
\node[yamlf] at (\codeX,  -7*\dy)  {\hspace*{3.5ex}max\_payload\_bytes: 4096};
\node[yamlf] at (\codeX,  -8*\dy)  {side\_effect\_level: read\_only};
\node[yamlf] at (\codeX,  -9*\dy)  {requires\_confirmation: false};
\node[yamlf] at (\codeX, -10*\dy)  {idempotent: true};
\node[yamlf] at (\codeX, -11*\dy)  {compatibility:};
\node[yamlf] at (\codeX, -12*\dy)  {\hspace*{3.5ex}allowed\_versions: [v1]};
\node[yamlf] at (\codeX, -13*\dy)  {drift\_policy:};
\node[yamlf] at (\codeX, -14*\dy)  {\hspace*{3.5ex}version\_mismatch: block};
\node[yamlf] at (\codeX, -15*\dy)  {\hspace*{3.5ex}audit\_class\_change: review, \ldots};
\node[yamlf] at (\codeX, -16*\dy)  {provenance:};
\node[yamlf] at (\codeX, -17*\dy)  {\hspace*{3.5ex}source\_path: \ldots/catalog.yaml};
\node[yamlf] at (\codeX, -18*\dy)  {\hspace*{3.5ex}openapi\_sha256: 0aaf9e9b\ldots};
\node[yamlf] at (\codeX, -19*\dy)  {annotations:};
\node[yamlf] at (\codeX, -20*\dy)  {\hspace*{3.5ex}readOnlyHint: true};
\node[yamlf] at (\codeX, -21*\dy)  {\hspace*{3.5ex}destructiveHint: false, \ldots};
\node[yamlf] at (\codeX, -22*\dy)  {input\_schema: \{type: object, \ldots\}};

\draw[color=tcgblue!70!black, line width=0.8pt]
  (\bracX, 0.10) -- (\bracX+\bw, 0.10)
  -- (\bracX+\bw, -3*\dy - 0.10) -- (\bracX, -3*\dy - 0.10);
\node[seclbl, color=tcgblue!70!black] at (\labX+\bw, -1.5*\dy) {Identity};

\draw[color=tcgblue!40!black, line width=0.8pt]
  (\bracX, -4*\dy + 0.10) -- (\bracX+\bw, -4*\dy + 0.10)
  -- (\bracX+\bw, -5*\dy - 0.10) -- (\bracX, -5*\dy - 0.10);
\node[seclbl, color=tcgblue!40!black] at (\labX+\bw, -4.5*\dy) {Access};

\draw[color=tcggreen!70!black, line width=0.8pt]
  (\bracX, -6*\dy + 0.10) -- (\bracX+\bw, -6*\dy + 0.10)
  -- (\bracX+\bw, -7*\dy - 0.10) -- (\bracX, -7*\dy - 0.10);
\node[seclbl, color=tcggreen!70!black] at (\labX+\bw, -6.5*\dy) {Limits};

\draw[color=tcgorange!75!black, line width=0.8pt]
  (\bracX, -8*\dy + 0.10) -- (\bracX+\bw, -8*\dy + 0.10)
  -- (\bracX+\bw, -10*\dy - 0.10) -- (\bracX, -10*\dy - 0.10);
\node[seclbl, color=tcgorange!75!black] at (\labX+\bw, -9*\dy) {Behavior};

\draw[color=tcgblue!50!black, line width=0.8pt]
  (\bracX, -11*\dy + 0.10) -- (\bracX+\bw, -11*\dy + 0.10)
  -- (\bracX+\bw, -12*\dy - 0.10) -- (\bracX, -12*\dy - 0.10);
\node[seclbl, color=tcgblue!50!black] at (\labX+\bw, -11.5*\dy) {Compatibility};

\draw[color=tcgred!70!black, line width=0.8pt]
  (\bracX, -13*\dy + 0.10) -- (\bracX+\bw, -13*\dy + 0.10)
  -- (\bracX+\bw, -15*\dy - 0.10) -- (\bracX, -15*\dy - 0.10);
\node[seclbl, color=tcgred!70!black] at (\labX+\bw, -14*\dy) {Drift policy};

\draw[color=tcgblue!85!black, line width=0.8pt]
  (\bracX, -16*\dy + 0.10) -- (\bracX+\bw, -16*\dy + 0.10)
  -- (\bracX+\bw, -18*\dy - 0.10) -- (\bracX, -18*\dy - 0.10);
\node[seclbl, color=tcgblue!85!black] at (\labX+\bw, -17*\dy) {Provenance};

\draw[color=tcgorange!50!black, line width=0.8pt]
  (\bracX, -19*\dy + 0.10) -- (\bracX+\bw, -19*\dy + 0.10)
  -- (\bracX+\bw, -21*\dy - 0.10) -- (\bracX, -21*\dy - 0.10);
\node[seclbl, color=tcgorange!50!black] at (\labX+\bw, -20*\dy) {MCP Annot.};

\draw[color=gray!65!black, line width=0.8pt]
  (\bracX, -22*\dy + 0.10) -- (\bracX+\bw, -22*\dy + 0.10)
  -- (\bracX+\bw, -22*\dy - 0.10) -- (\bracX, -22*\dy - 0.10);
\node[seclbl, color=gray!65!black] at (\labX+\bw, -22*\dy) {Input Schema};

\end{tikzpicture}
\caption{\textbf{Anatomy of a ToolContract.} Annotated excerpt from the contract YAML \texttt{catalog.search.v1.yaml} (under \texttt{gateway/src/\allowbreak{}main/\allowbreak{}resources/\allowbreak{}contracts/}). Colored brackets group the nine typed sections referenced throughout the paper. The \texttt{openapi\_sha256} field is verified at boot by Algorithm~\ref{alg:provenance}; the \texttt{annotations} block is field-aligned with MCP \texttt{ToolAnnotations}~\cite{mcp_spec}.}
\label{fig:contract}
\end{figure}

\subsection{Checks, decisions, and actions}

Table~\ref{tab:checks} summarizes the ten checks in evaluation order. A violation records its type, severity, expected value, and observed value. The final \texttt{decision} is the maximum under
\[\textsc{Block} > \textsc{Review} > \textsc{Warn} > \textsc{Allow}.\]
The last visited violation tied at that maximum supplies \texttt{reasonCode}; all violations remain available. For an enforcing host, BLOCK suppresses the call, REVIEW holds it for review, and WARN permits it with a warning. The absence of violations yields ALLOW. These are integration semantics; a decision response alone does not execute, suppress, or review a downstream call.

\begin{table*}[t]
\centering
\caption{Ten runtime checks in pipeline order. Defaults apply with the full configuration and default drift policy; the integrating host implements the indicated action.}
\label{tab:checks}
\small
\setlength{\tabcolsep}{4pt}
\renewcommand{\arraystretch}{1.08}
\begin{tabularx}{\textwidth}{@{}r p{0.13\textwidth} X l l@{}}
\toprule
\textbf{\#} & \textbf{Check} & \textbf{Violation condition} & \textbf{Emitted type} & \textbf{Default/action}\\
\midrule
1 & Identity & No contract resolves for the supplied ID. & \texttt{contract\_not\_found} & BLOCK / deny\\
2 & Version & Requested version is not in the allowed set. & \texttt{version\_mismatch} & BLOCK / deny\\
3 & Tool binding & Requested tool name differs from the bound name. & \texttt{tool\_mismatch} & BLOCK / deny\\
4 & Scope & A required scope is absent from supplied scopes. & \texttt{scope\_violation} & BLOCK / deny\\
5 & Input schema & Payload fails the compiled JSON Schema. & \texttt{schema\_violation} & BLOCK / deny\\
6 & Provenance & Provenance is present; cached validity is false. & \texttt{provenance\_mismatch} & BLOCK / deny\\
7 & Side effect & Non-null declared level exceeds the contract level. & \texttt{side\_effect\_change} & BLOCK / deny\\
8 & Confirmation & Contract requires it; supplied flag is explicitly false. & \texttt{confirmation\_change} & BLOCK / deny\\
9 & Payload limit & Serialized JSON UTF-8 bytes exceed the limit. & \texttt{payload\_limit\_violation} & WARN / permit\\
10 & Audit class & Non-null declared class is below the contract class. & \texttt{audit\_class\_change} & REVIEW / hold\\
\bottomrule
\end{tabularx}
\vspace{3pt}
\begin{minipage}{\textwidth}\footnotesize
Rows 1 and 6 start at hardcoded BLOCK; other severities use contract policy. Row 9 uses policy key \texttt{limit\_change}. \texttt{warn\_only} downgrades BLOCK/REVIEW to WARN, including rows 1 and 6. Version, scope, schema, and provenance flags control rows 2, 4, 5, and 6; the annotation flag controls rows 7--8. Rows 1, 3, 9, and 10 have no individual disable flag. No violation yields ALLOW.
\end{minipage}
\end{table*}

\subsection{Modes and per-call configuration}

\begin{figure}[t]
\centering
\begin{tikzpicture}[x=1cm,y=1cm,font=\footnotesize,
  box/.style={draw,rounded corners=2pt,align=center,inner sep=4pt,text width=7.15cm},
  half/.style={draw,rounded corners=2pt,align=center,inner sep=4pt,text width=3.28cm},
  arr/.style={-{Stealth[length=4pt]},semithick}]
\node[box,fill=tcglight] (input) at (0,0) {Request + mode + configuration};
\node[box,below=7pt of input,fill=tcglight!50] (presets) {\textbf{Five measured configurations}\\
\texttt{full}: all checks, normal severity\\
\texttt{no\_version} / \texttt{no\_provenance} / \texttt{no\_schema}:\\disable the named check\\
\texttt{warn\_only}: BLOCK / REVIEW $\rightarrow$ WARN};
\node[box,below=7pt of presets] (modes) {\textbf{Select evaluation path}\\
\texttt{NO\_GATE}: bypass checks, ALLOW\\
\texttt{SCHEMA\_ONLY}: schema check if enabled and resolved\\
\texttt{TOOL\_CONTRACT\_GATE}, \texttt{SHADOW\_GATE}:\\ten-check pipeline with configured flags};
\node[box,below=7pt of modes,fill=tcgblue!8] (decision) {Aggregate violations $\rightarrow$ \texttt{decision}\\BLOCK $>$ REVIEW $>$ WARN $>$ ALLOW};
\node[half,anchor=north east,fill=tcggreen!10] (normal) at (-0.13,-6.1) {\textbf{Non-shadow modes}\\\texttt{runtimeDecision}\\$=$ \texttt{decision}};
\node[half,anchor=north west,fill=tcggray!12] (shadow) at (0.13,-6.1) {\textbf{Shadow mode}\\\texttt{runtimeDecision}\\$=$ ALLOW};
\draw[arr] (input) -- (presets);
\draw[arr] (presets) -- (modes);
\draw[arr] (modes) -- (decision);
\draw[arr] (decision.south) -- ++(0,-0.15) -| (normal.north);
\draw[arr] (decision.south) -- ++(0,-0.15) -| (shadow.north);
\end{tikzpicture}
\caption{Modes and configurations are separate choices. NO\_GATE returns early; the remaining paths retain their decision for analysis. Schema-only respects \texttt{schema\_check} and \texttt{warn\_only}. Shadow preserves would-be severity but always permits runtime execution. Host enforcement and audit persistence are outside this evaluator.}
\label{fig:modes}
\end{figure}
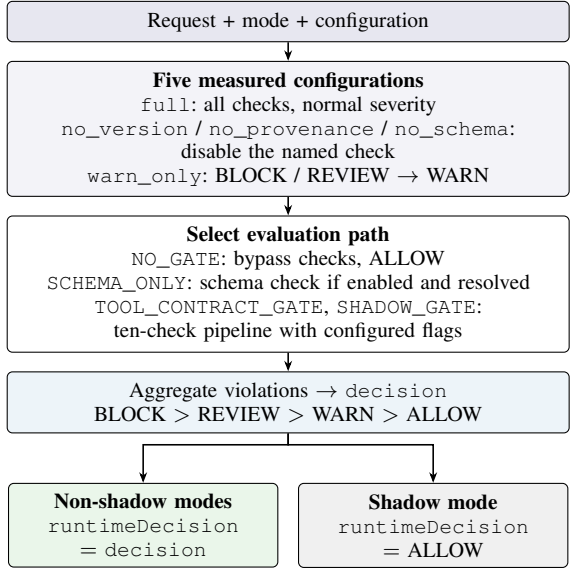

Figure~\ref{fig:modes} relates four modes to five measured configurations. \texttt{NO\_GATE} returns ALLOW immediately. \texttt{SCHEMA\_ONLY} validates the payload only when the contract resolves and schema checking is enabled. \texttt{TOOL\_CONTRACT\_GATE} evaluates all enabled checks and sets \texttt{runtimeDecision=decision}. \texttt{SHADOW\_GATE} evaluates the same checks but forces \texttt{runtimeDecision=ALLOW}, retaining the would-be decision.

\texttt{GateConfig} has five check flags (version, schema, provenance, scope, annotation) plus \texttt{warn\_only}. The annotation flag controls both side-effect and confirmation checks. The five measured presets are \texttt{full}, \texttt{no\_version}, \texttt{no\_provenance}, \texttt{no\_schema}, and \texttt{warn\_only}; scope and annotation ablations are available but not measured here. \texttt{warn\_only} changes BLOCK and REVIEW violations to WARN at construction, including early-return paths. Production use requires a trusted host to choose modes/flags; the experimental endpoint does not authenticate their source.

\section{Algorithms}
\label{sec:algorithms}

Algorithm~\ref{alg:gate} evaluates predicates in Table~\ref{tab:checks} order. Only baseline modes and unresolved-contract handling return early. The helper \textsc{Add} creates a violation using the policy key associated with its emitted type, then applies \texttt{warn\_only}. Contract absence and invalid provenance start at BLOCK independently of contract policy. Payload overflow emits \texttt{payload\_limit\_violation} but looks up severity using \texttt{limit\_change}. \textsc{Finish} aggregates severity and attaches the mode-specific runtime decision.

\begin{algorithm}[t]
\caption{Runtime contract evaluation}
\label{alg:gate}
\small
\begin{algorithmic}[1]
\Require Input $x$; configuration $g$ (full if absent)
\Ensure Decision and mode-specific runtime decision
\State $V\gets[\,]$ \Comment{Ordered violation list}
\If{$x.\mathrm{mode}=\texttt{NO\_GATE}$}
 \State \Return \Call{Finish}{$V,x$}
\EndIf
\State $c\gets$ \Call{ResolveById}{$x.\mathrm{id}$}
\If{$x.\mathrm{mode}=\texttt{SCHEMA\_ONLY}$}
 \If{$c\ne\bot$ and $g.\mathrm{schemaCheck}$}
  \State \Call{CheckSchema}{$c,x,g,V$}
 \EndIf
 \State \Return \Call{Finish}{$V,x$}
\EndIf
\If{$c=\bot$}
 \State \Call{Add}{$V,\mathrm{contract\_not\_found},\mathrm{BLOCK},g$}
 \State \Return \Call{Finish}{$V,x$}
\EndIf
\For{$j=2,\ldots,10$ in Table~\ref{tab:checks} order}
 \If{check $j$ is enabled and its violation predicate holds}
  \State $t\gets$ emitted violation type for $j$
  \State $s\gets\mathrm{BLOCK}$ if $j=6$, else contract policy
  \State \Call{Add}{$V,t,s,g$}
 \EndIf
\EndFor
\State \Return \Call{Finish}{$V,x$}
\end{algorithmic}
\vspace{3pt}
\footnotesize
\textsc{Add} attaches the expected/observed values for the predicate; if $g.\mathrm{warnOnly}$ and $s\in\{\mathrm{BLOCK},\mathrm{REVIEW}\}$, it changes $s$ to WARN before appending. \textsc{CheckSchema} uses the same helper for a schema failure. \textsc{Finish} collapses by maximum severity (last equal-severity reason wins), then forces runtime ALLOW only for SHADOW\_GATE. Otherwise runtime equals decision. Null guards and flag mappings are given in Table~\ref{tab:checks}.
\end{algorithm}

\begin{algorithm}[t]
\caption{Boot-time provenance and schema initialization}
\label{alg:provenance}
\small
\begin{algorithmic}[1]
\Require YAML contract resources $\mathcal R$
\For{$r\in\mathcal R$}
 \State $c\gets$ \Call{FromYaml}{$r$}
 \If{$c.\mathrm{provenance}\ne\bot$}
  \State $b\gets$ \Call{ReadFileOrClasspath}{$c.\mathrm{sourcePath}$}
  \State $c.\mathrm{valid}\gets(b\ne\bot)$ and
  \Statex \hspace{\algorithmicindent}\hspace{\algorithmicindent}\hspace{\algorithmicindent}$\bigl(\mathrm{SHA256}(b)=c.\mathrm{storedDigest}\bigr)$
 \EndIf
 \State Register $c$ and initialize its current-version entry
 \State Compile and cache $c$'s input schema
\EndFor
\end{algorithmic}
\end{algorithm}

Algorithm~\ref{alg:provenance} runs at startup. If provenance is present, the registry reads the referenced file from the working directory or classpath and compares its SHA-256 with the stored digest. Missing files or mismatching digests set the cached validity bit to false. The contract remains registered; an enabled full-pipeline provenance check subsequently emits a violation. The verifier does not compare an OpenAPI operation's semantics with a contract schema or a deployed backend.

\section{Implementation}
\label{sec:impl}

The artifact uses Java 21, Spring Boot 3.3.4, and networknt JSON Schema Validator 1.5.2 with JSON Schema 2020-12. At boot, YAML contracts load from the classpath; schemas compile once, and provenance validity is cached. Seven OpenAPI 3.1 files support twelve enterprise contracts across catalog, inventory, profile, ticket, notifications, audit, and feature-flag services. Three additional legacy contracts support the demo and have no provenance block; they are excluded from the 45-case benchmark.

\texttt{POST /v1/decide} accepts the gate input and returns a \texttt{Decision}, including violations, reason, runtime decision, and server-measured \texttt{gateLatencyMs}. Well-formed evaluations return HTTP 200 even for BLOCK: the consumer must inspect the body. The endpoint logs a short decision summary, not a persistent structured audit record, and does not forward calls or implement a human-review channel. The experiment supplies scopes and metadata directly, rather than extracting them from authenticated production traffic.

The legacy \texttt{/v1/invoke} endpoint validates JSON Web Tokens (JWTs) and performs early contract, version, tool, scope, and schema checks, then returns a stub result. Its later call to the new evaluator records the additional gate result; it does not condition the stub response on that result. Thus the benchmark evaluates decision coverage, not complete downstream enforcement.

\subsection{Blocked-invocation audit example}

Listing~\ref{lst:audit} shows captured output from the existing legacy mismatch demo. The request references \texttt{metrics@v1}, which binds \texttt{prom\_query}, but asks for \texttt{vault.read\_secret}. The tool-name check rejects the invocation with HTTP 403 before reaching the stub. The diagnostic gives expected and requested names; the audit line records the denial category, and the response carries a trace identifier for correlation. This operational example uses the legacy logger's \texttt{deny} terminology, distinct from the gate-only endpoint's structured BLOCK decision. It demonstrates diagnostic output, not durable or tamper-evident audit storage.

\begin{lstlisting}[caption={Captured denial: log and HTTP response excerpts, reflowed for readability.},label={lst:audit},basicstyle=\ttfamily\footnotesize]
[gateway] tool.MISMATCH
 requested=vault.read_secret
 contract.tool=prom_query
[audit] write class=sensitive
 decision=deny reason=tool_mismatch

HTTP 403
{"error": "contract_tool_mismatch",
 "trace_id": "b3db08ca"}
\end{lstlisting}

The source and archived results are available at \url{https://github.com/sachinkg12/toolcontractgate}, commit \texttt{3eb1f67}. The camera-ready review preserves this implementation baseline.

\section{Evaluation}
\label{sec:eval}

\subsection{Case design and metrics}

The benchmark tests deterministic predicate coverage rather than population-level drift prevalence. Its 45 YAML cases comprise 35 unsafe cases across twelve drift classes, five happy paths, and five safe-compatibility variants. Safe variants include extra scopes, voluntary confirmation, and a stronger declared audit class. Table~\ref{tab:fault-detection} gives category counts. Large-payload placeholders are expanded by the runner; enum and required-field cases exercise schema validation. Wrong-operation-binding cases change the supplied tool name; they do not test unchanged-name backend substitution.

Each case runs under four modes and five presets, yielding 900 records. We distinguish three metrics. \emph{Detected} means a non-ALLOW evaluator decision on an unsafe case, including WARN. The \emph{B/R count} includes evaluator BLOCK or REVIEW decisions on unsafe cases; in shadow mode these are would-block/review decisions. The \emph{false-block count} includes safe cases from either bucket that receive runtime BLOCK or REVIEW. These metrics are computed from the CSV fields; the runner's printed counters omit safe-change cases from its false-block denominator and therefore are not used as the metric definition.

\begin{table}[t]
\centering
\caption{Detection rate by fault type, ablation $=$ \texttt{full}. A case is detected when the gate returns any non-ALLOW decision; the two safe buckets report runtime BLOCK/REVIEW counts, which are zero in all modes. TCG abbreviates the \texttt{TOOL\_CONTRACT\_GATE} mode (not the system \tcg{}).}
\label{tab:fault-detection}
\small
\setlength{\tabcolsep}{3.4pt}
\begin{tabular}{lrcccc}
\toprule
\textbf{Fault type} & \textbf{N} & \textbf{NG} & \textbf{SO} & \textbf{SG} & \textbf{TCG} \\
\midrule
\multicolumn{6}{l}{\emph{Unsafe drift (35 cases)}}\\
tool\_mismatch          & 4 & 0/4 & 0/4 & 4/4 & 4/4 \\
version\_mismatch       & 3 & 0/3 & 0/3 & 3/3 & 3/3 \\
schema\_violation       & 4 & 0/4 & 4/4 & 4/4 & 4/4 \\
missing\_scope          & 4 & 0/4 & 0/4 & 4/4 & 4/4 \\
side\_effect\_change    & 3 & 0/3 & 0/3 & 3/3 & 3/3 \\
confirmation\_change    & 3 & 0/3 & 0/3 & 3/3 & 3/3 \\
payload\_limit          & 3 & 0/3 & 0/3 & 3/3 & 3/3 \\
audit\_class\_change    & 3 & 0/3 & 0/3 & 3/3 & 3/3 \\
destr.\ w/o conf.       & 2 & 0/2 & 0/2 & 2/2 & 2/2 \\
enum\_drift             & 2 & 0/2 & 2/2 & 2/2 & 2/2 \\
required\_field\_added  & 2 & 0/2 & 2/2 & 2/2 & 2/2 \\
wrong\_op.\ binding     & 2 & 0/2 & 0/2 & 2/2 & 2/2 \\
\addlinespace[2pt]
\textbf{Drift subtotal} & \textbf{35} & \textbf{0/35} & \textbf{8/35} & \textbf{35/35} & \textbf{35/35} \\
\midrule
\multicolumn{6}{l}{\emph{Safe (10 cases): runtime BLOCK/REVIEW}}\\
none (happy)            & 5 & 0/5 & 0/5 & 0/5 & 0/5 \\
safe\_change            & 5 & 0/5 & 0/5 & 0/5 & 0/5 \\
\addlinespace[2pt]
\textbf{Safe subtotal}  & \textbf{10} & \textbf{0/10} & \textbf{0/10} & \textbf{0/10} & \textbf{0/10} \\
\bottomrule
\multicolumn{6}{l}{\textsmaller{NG = \texttt{NO\_GATE}; SO = \texttt{SCHEMA\_ONLY}; SG = \texttt{SHADOW\_GATE}. Names}}\\
\multicolumn{6}{l}{\textsmaller{abbreviated for layout; full identifiers in \texttt{experiments/cases.yaml}.}}
\end{tabular}
\end{table}

\subsection{Detection and ablation results}

The full gate detects 35/35 unsafe cases: 29 BLOCK, three REVIEW, and three WARN decisions. The three warnings are payload-limit cases under the default policy. All ten safe cases return ALLOW. The schema-only baseline detects 8/35 unsafe cases; the no-gate baseline detects none. These outcomes establish coverage for the supplied cases, not a false-positive rate for production workloads.

Table~\ref{tab:matrix} reports all twenty mode/preset combinations. Disabling version checking removes three detections; disabling schema checking removes eight. \texttt{warn\_only} retains all 35 detections while eliminating BLOCK/REVIEW decisions. Disabling provenance has no effect because no provenance-mismatch case occurs in the 45-case set; startup integrity is checked separately.

\begin{table*}[t]
\centering
\caption{All configurations, 45 cases each. Detected and B/R count evaluator non-ALLOW and BLOCK/REVIEW on the 35 unsafe cases. Runtime B/R counts runtime BLOCK/REVIEW on unsafe cases; these are returned decisions, not measured downstream effects. False blocks count runtime BLOCK/REVIEW on the 10 safe cases. Latencies are archived server-side integer-millisecond measurements.}
\label{tab:matrix}
\small
\setlength{\tabcolsep}{5pt}
\begin{tabular}{@{}llrrrrrr@{}}
\toprule
\textbf{Mode} & \textbf{Preset} & \textbf{Detected} & \textbf{B/R} & \textbf{Runtime B/R} & \textbf{False blocks} & \textbf{Mean (ms)} & \textbf{p95 (ms)}\\
\midrule
\texttt{NO\_GATE} & full & 0 & 0 & 0 & 0 & 0.02 & 0 \\
\texttt{NO\_GATE} & no\_version & 0 & 0 & 0 & 0 & 0.00 & 0 \\
\texttt{NO\_GATE} & no\_provenance & 0 & 0 & 0 & 0 & 0.00 & 0 \\
\texttt{NO\_GATE} & no\_schema & 0 & 0 & 0 & 0 & 0.00 & 0 \\
\texttt{NO\_GATE} & warn\_only & 0 & 0 & 0 & 0 & 0.00 & 0 \\
\addlinespace[2pt]
\texttt{SCHEMA\_ONLY} & full & 8 & 8 & 8 & 0 & 0.69 & 4 \\
\texttt{SCHEMA\_ONLY} & no\_version & 8 & 8 & 8 & 0 & 0.02 & 0 \\
\texttt{SCHEMA\_ONLY} & no\_provenance & 8 & 8 & 8 & 0 & 0.00 & 0 \\
\texttt{SCHEMA\_ONLY} & no\_schema & 0 & 0 & 0 & 0 & 0.00 & 0 \\
\texttt{SCHEMA\_ONLY} & warn\_only & 8 & 0 & 0 & 0 & 0.00 & 0 \\
\addlinespace[2pt]
\texttt{TOOL\_CONTRACT\_GATE} & full & 35 & 32 & 32 & 0 & 0.18 & 2 \\
\texttt{TOOL\_CONTRACT\_GATE} & no\_version & 32 & 29 & 29 & 0 & 0.20 & 0 \\
\texttt{TOOL\_CONTRACT\_GATE} & no\_provenance & 35 & 32 & 32 & 0 & 0.51 & 0 \\
\texttt{TOOL\_CONTRACT\_GATE} & no\_schema & 27 & 24 & 24 & 0 & 0.09 & 0 \\
\texttt{TOOL\_CONTRACT\_GATE} & warn\_only & 35 & 0 & 0 & 0 & 0.00 & 0 \\
\addlinespace[2pt]
\texttt{SHADOW\_GATE} & full & 35 & 32 & 0 & 0 & 0.02 & 0 \\
\texttt{SHADOW\_GATE} & no\_version & 32 & 29 & 0 & 0 & 0.09 & 0 \\
\texttt{SHADOW\_GATE} & no\_provenance & 35 & 32 & 0 & 0 & 0.04 & 0 \\
\texttt{SHADOW\_GATE} & no\_schema & 27 & 24 & 0 & 0 & 0.00 & 0 \\
\texttt{SHADOW\_GATE} & warn\_only & 35 & 0 & 0 & 0 & 0.02 & 0 \\
\addlinespace[2pt]\bottomrule
\end{tabular}
\end{table*}

Shadow mode matches the full gate's evaluator outcomes, including 32 BLOCK/REVIEW decisions under \texttt{full}, while all 45 runtime decisions remain ALLOW. Consequently its runtime false blocks are zero by construction, and its B/R count does not represent suppressed calls. For the full gate, the runtime decision equals the evaluator decision, but actual suppression still depends on the host.

The archived CSV also contains a \texttt{correct} field. Its schema-only oracle ignores ablation settings, so \texttt{no\_schema} and \texttt{warn\_only} each show 37/45 correct despite the expected behavior of their eight affected schema cases. Table~\ref{tab:matrix} reports raw outcome counts rather than treating these oracle mismatches as unexpected gate failures.

\subsection{Latency and provenance validation}
\label{sec:provenance-eval}

The archived run gives full-gate mean latency 0.18\,ms and p95 2\,ms; the largest p95 among its buckets is 4\,ms. These are server-side evaluation times from one sequential run on a developer machine. The timer truncates each duration to integer milliseconds, so zero denotes a duration below 1\,ms and the mean averages quantized observations. There is no controlled warm-up, repeated-trial confidence interval, concurrency study, or measured production RPC comparison. Differences between full and shadow timing should not be interpreted as an intrinsic speed advantage, since they share the pipeline.

The stored digests in all twelve enterprise contracts match the seven approved files. An isolated integrity exercise confirmed that editing an OpenAPI file after startup leaves decisions unchanged; restarting the gateway marks affected contracts invalid and returns \texttt{provenance\_mismatch} under the full configuration. Matching-hash, mismatching-hash, and missing-file unit tests cover the verifier. These checks establish byte-integrity behavior at load time, not detection of live semantic changes.

\subsection{Reproducibility}

A clean build uses \texttt{mvn package} from the repository root and runs 29 unit tests. Start the gateway from that root so relative OpenAPI paths resolve. Then invoke the runner with all five presets explicitly, as in Listing~\ref{lst:reproduce}. Use a fresh output path to preserve the archived CSV. The convenience wrapper at the pinned commit omits the preset argument and therefore defaults to \texttt{full}, producing 180 evaluations; the explicit invocation produces the complete 900-record matrix.

\begin{lstlisting}[caption={Complete evaluation from the repository root after building.},label={lst:reproduce},basicstyle=\ttfamily\scriptsize]
java -jar gateway/target/gateway-0.1.0.jar
# In a second terminal:
ABL=full,no_version,no_provenance,no_schema,warn_only
java -jar agent-runner/target/agent-runner.jar \
  experiments experiments/cases.yaml \
  /tmp/tcg-camera-ready-results.csv "$ABL"
\end{lstlisting}

A fresh camera-ready reproduction at the pinned commit passed all 29 tests and reproduced the decision, runtime-decision, reason-code, and oracle-correctness fields in all 900 archived records. Timing is environment-dependent; the paper retains the archived latency values rather than mixing timings from different runs.

\section{Discussion}

\subsection{Security role and trust boundary}

Workload identity (e.g., SPIFFE~\cite{spiffe_spec}), delegated authorization, and contract admission answer different questions: who is calling, what authority is conveyed, and whether the supplied invocation meets the approved contract. They can be composed, but the artifact does not implement their complete integration. Contract authoring and evaluation provide concrete governance, measurement, and management mechanisms in the spirit of NIST AI RMF~\cite{nist_airmf}. Scope and confirmation checks can support controls on excessive agency~\cite{owasp_llm_top10}, provided the host authenticates metadata and enforces decisions.

The trusted base includes the gateway, contracts, approved artifact store, and the integrating host. An attacker able to change both the contract digest and the source file can defeat the digest comparison. The SHA-256 linkage is compatible with artifact identification in in-toto~\cite{intoto_spec} and SLSA~\cite{slsa_spec}, but this prototype does not produce or verify those attestations. Contracts are manually seeded; automatic, reviewed contract derivation remains future work.

Modes, ablation flags, scopes, and optional behavior fields must come from a trusted integration. The experimental endpoint does not enforce that boundary. Missing side-effect or audit-class metadata skips the respective comparison; a missing confirmation flag is not an explicit false value. A boolean declaration does not establish that a human approved the call. Legacy contracts without provenance skip that check. Production integration must define required metadata, trusted policy selection, review routing, and audit retention explicitly.

\subsection{Semantic drift}
\label{sec:limitations}

ToolContractGate detects discrepancies exposed by its predicates, including mismatched names, invalid inputs, and declared side-effect escalation. It cannot prove that the actual operation matches its description. For example, a tool may keep its name, input schema, and read-only declaration while its backend begins writing records; that invocation can pass. The stored \texttt{operation\_id} is metadata, not a verified binding to the running handler. Likewise, a payload can satisfy the schema while expressing undesirable but permitted behavior. Behavioral probes, reviewed contract regeneration, provider tests, and execution monitoring are complementary future controls; the present benchmark does not evaluate them.

\subsection{Dynamic provenance verification}

Startup hashing creates a snapshot. An artifact edited after startup remains unchecked until another load or restart; the prototype exposes no automatic refresh hook. Rehashing per call would reduce the stale interval but add file access and hashing cost, and would still require binding the checked version to the operation actually invoked to avoid time-of-check/time-of-use races. Periodic or event-triggered verification amortizes cost but leaves a bounded or implementation-dependent stale window. An alternative is immutable versioned artifacts with atomic replacement of the approved contract, digest, and compiled schema. A dynamic design must define failure handling, rollback, and the trusted source of new digests. These alternatives are proposed extensions; their latency and integrity tradeoffs have not been measured here.

\subsection{External validity and deployment scale}

The 900 records repeat 45 constructed cases across configurations; they are not 900 independent production workloads. Ten safe examples do not characterize rare false blocks, changing usage patterns, or multi-tenant traffic. No live LLM/MCP agent traffic, throughput, concurrent deployment, or human-review workload is evaluated.

A production study should begin with authorized, redacted tool traces, independently reviewed labels, contract/version history, and representative safe and drift cases. Replay would measure detection and false-block rates; subsequent shadow deployment would measure would-block frequency and operator review volume. A separate scale experiment should vary registry size, schema complexity, payload size, and concurrency, recording hardware/JDK, warm-up, repeated trials, throughput, memory, errors, and p50/p95/p99 gate and end-to-end latency. Synthetic load tests would characterize scaling but would not substitute for production validity. Neither study is claimed as completed in this paper.

\section{Conclusion}

ToolContractGate combines versioned trusted contracts, boot-time OpenAPI integrity verification, and a deterministic ten-check evaluator for agent tool requests. Four modes and five evaluation configurations expose decision coverage and shadow behavior. On 35 constructed unsafe cases, the full evaluator detects every fault, producing 29 BLOCK, three REVIEW, and three WARN decisions, while allowing all ten safe cases. The open artifact and explicit reproduction command support inspection of all 900 evaluations. The demonstrated contribution is contract-decision evaluation; reliable deployment also requires trustworthy inputs and a host that enforces results. Future work addresses automated contract derivation, semantic behavior validation, dynamic provenance refresh, and production and scale evaluation.

\section*{Acknowledgment}

The author thanks the maintainers of Spring Boot, networknt JSON Schema Validator, Jackson, and jjwt, and the contributors to the MCP and SPIFFE specifications.

\newpage
\bibliographystyle{IEEEtran}
\bibliography{references}
\end{document}